\documentclass[hyper]{JHEP3}
\usepackage{graphicx}
\usepackage{amsfonts}
\usepackage{amssymb}
\usepackage{amsbsy}
\usepackage{amsmath}
\usepackage{array}

\newcommand*{\di}{\partial}
\newcommand{\be}{\begin{equation}}
\newcommand{\ee}{\end{equation}}
\newcommand{\bea}{\begin{eqnarray}}
\newcommand{\eea}{\end{eqnarray}}
\newcommand{\none}{\nonumber \\}
\newcommand{\ring}{\mathrm{ring}}

\title{Tomimatsu-Sato geometries, holography and quantum gravity}

\author{Jack Gegenberg, Haitao Liu, Sanjeev S.~Seahra, Benjamin K.\ Tippett \\ Department of Mathematics and Statistics \\ University of New Brunswick \\ Fredericton, NB  E3B 5A3 \\ Canada} 

\abstract{We analyze the $\delta=2$ Tomimatsu-Sato spacetime in the context of the proposed Kerr/CFT correspondence.  This 4-dimensional vacuum spacetime is asymptotically flat and has a well-defined ADM mass and angular momentum, but also involves several exotic features including a naked ring singularity, and two disjoint Killing horizons separated by a region with closed timelike curves and a rod-like conical singularity.  We demonstrate that the near horizon geometry belongs to a general class of Ricci-flat metrics with $SL(2,\mathbb{R})\times U(1)$ symmetry that includes both the extremal Kerr and extremal Kerr-bolt geometries.  We calculate the central charge and temperature for the CFT dual to this spacetime and confirm the Cardy formula reproduces the Bekenstein-Hawking entropy.  We find that all of the basic parameters of the dual CFT are most naturally expressed in terms of charges defined intrinsically on the horizon, which are distinct from the ADM charges in this geometry.}

\begin{document}

\section{Introduction}

There is a broad consensus in gravitational physics that the endstate of the collapse of uncharged matter is the Kerr black hole.  This belief is implicitly based on the cosmic censorship conjecture \cite{Wald:1997wa}, which states that curvature singularities in general relativity are necessarily hidden behind event horizons.  Since the positive-mass Kerr black hole is the only known solution of the vacuum field equations that is stationary, axisymmetric, stable, asymptotically flat and free from curvature singularities in causal contact with null infinity, it is the only viable candidate for the final configuration of collapsed matter consistent with the cosmic censorship conjecture.

However, it is worthwhile noting that cosmic censorship has yet to be proved, and there are infinitely many other axisymmetric, stationary, and asymptotically flat solutions of the vacuum Einstein field equations involving naked singularities.  It is interesting to study these solutions as possible alternatives to the Kerr metric for the description of the gravitational field around a compact astrophysical body with given mass and angular momentum.  It is also interesting to consider these solutions from a completely different perspective: namely, in the context of quantum gravity.  In recent years, there has been much activity in obtaining dual descriptions of classical black hole spacetimes by conformal field theories (CFTs) living on their boundaries.  A natural question is: given that the Einstein field equations admit non-black hole solutions describing collapsed objects, is it possible to describe such spacetimes using a dual CFT?

To attempt to address this question, we will focus on the Tomimatsu-Sato geometries \cite{PhysRevLett.29.1344,PTP.50.95}, which are a class of vacuum solutions in general relativity labelled by a parameter $\delta$.  For $\delta = 1$, the Tomimatsu-Sato spacetimes reduces down to the Kerr solution, but for $\delta \ne 1$ they involve naked singularities.  In all cases, the solutions are asymptotically flat, axisymmetric and stationary.  The algebraic complexity of these solutions increases rapidly with $\delta$, so for this reason the $\delta = 2$ case (which we denote by TS2) has been most intensively studied in the literature  \cite{Gibbons:1973aq,ernst:1091,Hikida:2003ar,Kodama:2003ch}.  For this choice, the spacetime contains a ring singularity on the boundary of an ergoregion.  At the center of the geometry is a rod-like conical singularity which carries all of the spacetime's gravitational mass, and is surrounded by a region containing closed timelike curves.  To date, it is not clear whether or not physical matter can collapse to a TS2 geometry, and if the TS2 geometry is stable.

In this paper, we calculate the properties of a candidate CFT dual to the TS2 geometry using tools from the recently proposed Kerr/CFT correspondence \cite{Guica:2008mu}, which is in turn inspired by the AdS/CFT correspondence \cite{Aharony:1999ti}.  The latter approach has been particularly successful in giving a quantum mechanical description of the thermodynamics of a large variety of extremal black holes whose asymptotic region is a space of constant negative curvature (for which supersymmetry allows for a continuation from high to low energy scales).  The essence of the calculation is the identification of a CFT living on the boundary of the spacetime that is in some sense ``dual'' to the black hole horizon.


However, the fact that this duality was initially restricted to asymptotically anti-de Sitter (AdS) black holes was not entirely satisfying, given the mounting physical evidence that the large scale geometry of our universe does not have an overall negative curvature.  The discovery that extremal Kerr black holes are dual to a CFT \cite{Guica:2008mu,Hartman:2008pb,Compere:2009dp} partially overcomes this problem.  This Kerr/CFT correspondence depends on the fact that the near horizon geometry of the extremal Kerr metric (NHEK) is a warped product of two dimensional AdS and $S^2$ \cite{Bardeen:1999px}.  The asymptotic region of the NHEK metric supports a CFT, enough of whose properties are known to construct thermodynamics which describe the original extremal Kerr black hole.  One problem is that, unlike the original AdS/CFT correspondence, the boundary CFT lives on a spacetime which cannot be identified with the boundary of the asymptotically {\it flat} region of the original Kerr black hole;  rather the CFT lives on the boundary of the NHEK geometry, which is {\it not} asymptotically flat.  In spite of this caveat, the program of enlarging the class of geometries which have a dual CFT description is important for our eventual understanding of quantum gravity, and should be pursued.  In this spirit, the original Kerr/CFT correspondence has been extended in a number of directions: including the CFT description of near-extremal Kerr black holes obtained by Castro \& Larsen \cite{Castro:2010fd} and Kerr-Bolt spacetimes obtained by Ghezelbash \cite{Ghezelbash:2009gy}.

In the current work, we obtain the further extension of the Kerr/CFT correspondence to non-black hole spacetimes, namely the TS2 geometry.  Our treatment is based on the discovery by Kodama and Hikida that, despite the existence of the naked ring singularity, the TS2 geometry does indeed contain a pair of disjoint Killing horizons located at each tip of the rod at the center of the spacetime \cite{Hikida:2003ar,Kodama:2003ch}.  Each horizon has isometry group $SL(2,\mathbb{R})\times U(1)$ and is extremal in the sense that it has zero surface gravity (i.e.\ it is degenerate).  These properties are shared by the horizons in extremal Kerr and extremal Kerr-Bolt, so it is reasonable to postulate that the TS2 geometry also admits a dual CFT description.  We demonstrate that this is indeed that case by first showing that the near horizon geometry in the TS2 solution belongs to a general class of Ricci flat solutions we call the {\it near horizon extremal spinning} (NHES) metrics.  These represent the most general vacuum metrics describing the geometry near a non-toriodal, axisymmetric, extremal horizon in general relativity \cite{Kunduri:2008rs}, and include the near horizon Kerr and Kerr-Bolt solutions.  We then proceed to obtain a dual CFT description for entire class of NHES geometries by computing the central charge from asymptotic symmetries \cite{Guica:2008mu}, and the temperature from analyzing the hidden symmetries of the scalar wave equation \cite{Castro:2010fd,Chen:2010fr}.  When inserted into the Cardy formula, our derived central charge and temperature successfully reproduce the Bekenstein-Hawking entropy of the horizon. 

We find that the basic parameters of the dual CFT are most naturally expressed in terms of the charges defined {\it intrinsically} on the horizon, rather than the ADM charges associated with the symmetries of the parent spacetime.  For example, in the TS2 geometry the central charge is proportional to the angular momentum of one of the horizons as opposed to the ADM angular momentum of the entire spacetime.  This seems to suggest that the CFT we discuss is really dual to the TS2 Killing horizon(s) and not the entire spacetime---indeed, the CFT we derive shows no awareness of the closed timelike curve region, ring singularity, or other pathologies of the full TS2 geometry.

The plan of the paper is as follows:  In \S\ref{sec:classical}, we give a self-contained analysis of the basic properties of the TS2 metrics, including the ring curvature singularity, the existence of stationary observers, and the distribution of mass and angular momentum, the near horizon geometry and the nature of the rod singularity.  In \S\ref{sec:CFT}, we calculate the central charge and temperature of the dual CFT and confirm that the macroscopic and microscopic entropies agree.  In \S\ref{sec:discussion}, we conclude with a discussion of the the relevant classical and quantum properties of the TS2 geometry.  In the appendix \S\ref{app:NHES}, we briefly discuss the uniqueness of the NHES geometries.

\section{Classical features of the $\delta=2$ Tomimatsu-Sato spacetime}\label{sec:classical}

In this section we introduce and analyze the $\delta = 2$ Tomimatsu-Sato (TS2) spacetime $(\mathcal{M},g)$ \cite{PhysRevLett.29.1344,PTP.50.95}.  The discussion parallels and extends several previous studies  \cite{Gibbons:1973aq,ernst:1091,Hikida:2003ar,Kodama:2003ch}.

\subsection{Line element and coordinate systems}

In prolate spherical coordinates $(x,y)$, the TS2 line element is
\begin{equation}\label{eq:line element}
ds^2=-f(d\tilde{t}-\omega d\phi)^2+ \frac{\sigma^2}{f} \left[E\left(\frac{dx^2}{x^2-1}+\frac{dy^2}{1-y^2}\right)+(x^2-1)(1-y^2)d\phi^2\right].
\end{equation}
Here we have
\begin{gather}\nonumber
f = f(x,y) \equiv \frac{A(x,y)}{B(x,y)}, \quad E = E(x,y) \equiv \frac{A(x,y)}{p^4(x^2-y^2)^3}, \\  \omega = \omega(x,y) \equiv \frac{4\sigma qC(x,y)}{pA(x,y)}(1-y^2),
\end{gather}
where
\begin{eqnarray}
A(x,y) & \equiv & p^4(x^2-1)^4+q^4(1-y^2)^4 
 -2p^2q^2(x^2-1)(1-y^2) \nonumber \\ & & \times \left[2(x^2-1)^2+2(1-y^2)^2+3(x^2-1)(1-y^2)\right];\none
B(x,y) & \equiv & \left[p^2(x^4-1)-q^2(1-y^4)+2px(x^2-1)\right]^2
+ \nonumber \\ & & 4q^2y^2\left[px(x^2-1)+(px+1)(1-y^2)\right]^2;\none
C(x,y) & \equiv & q^2(px+1)(1-y^2)^3 -p^3x(x^2-1)\left[2(x^4-1)+(x^2+3)(1-y^2)\right]  \none & &
-p^2(x^2-1)\left[4x^2(x^2-1)  +(3x^2+1)(1-y^2)\right].
\end{eqnarray}
Also, $\sigma$ is a constant with dimensions of length, while $p$ and $q$ are dimensionless constants.  This metric is Ricci-flat ($R_{\alpha\beta}=0$) for any $\sigma$ if we impose
\begin{equation}
	p^{2} + q^{2} = 1.
\end{equation}

The prolate spheroidal coordinates $(x,y)$ appearing in (\ref{eq:line element}) are related to the quasi-cylindrical coordinates $(\rho,z)$ by
\begin{equation}
\rho^2 = \sigma^{2}(x^2-1)(1-y^2), \quad z = \sigma xy,
\end{equation}
where $\sigma$ is a constant length parameter.  Below, we will also make use of the bipolar coordinate system $(\tilde{r},\theta)$, defined by
\begin{equation}\label{eq:bipolar}
	x^{2} = \frac{\sigma}{\sigma - \tilde{r} \cos^{2}(\theta/2)}, \quad
	y^{2} = \frac{\sigma-\tilde{r}}{\sigma - \tilde{r} \cos^{2}(\theta/2)}.
\end{equation}
We take the ranges of the various coordinates to be
\begin{align}\nonumber
	\rho & \in [0,\infty), & z & \in (-\infty,\infty), & x & \in [1,\infty), & y & \in [-1,1], \\ \tilde{r} & \in [0,\sigma], & \theta & \in [0,\pi], & \phi & \in [0,2\pi].
\end{align}
The coordinate traces of the $(x,y)$ and $(\tilde{r},\theta)$ coordinates in the $(\rho,z)$ plane are illustrated in Fig.\  \ref{fig:coords}.  Note that the bipolar coordinates only cover the upper half of the $(\rho,z)$ plane if we solve for $(x,y)$ by taking the positive square roots of equations (\ref{eq:bipolar}).  Also on this figure are shown special locations,
\begin{subequations}
\begin{align}
	\mathcal{S} & = \{ p \in \mathcal{M} | x=1,\,|y| < 1 \text{ or } \rho = 0,\, |z| < \sigma  \text{ or } \theta = \pi\}, \\ \mathcal{H}^\pm & = \{p \in \mathcal{M} | x=1, \, y = \pm 1 \text{ or } \rho = 0 ,\, z = \pm \sigma \text{ or } r = 0 \}, \end{align}
\end{subequations}
that we will study in detail below.
\FIGURE[t]{
\includegraphics[width=5.5in]{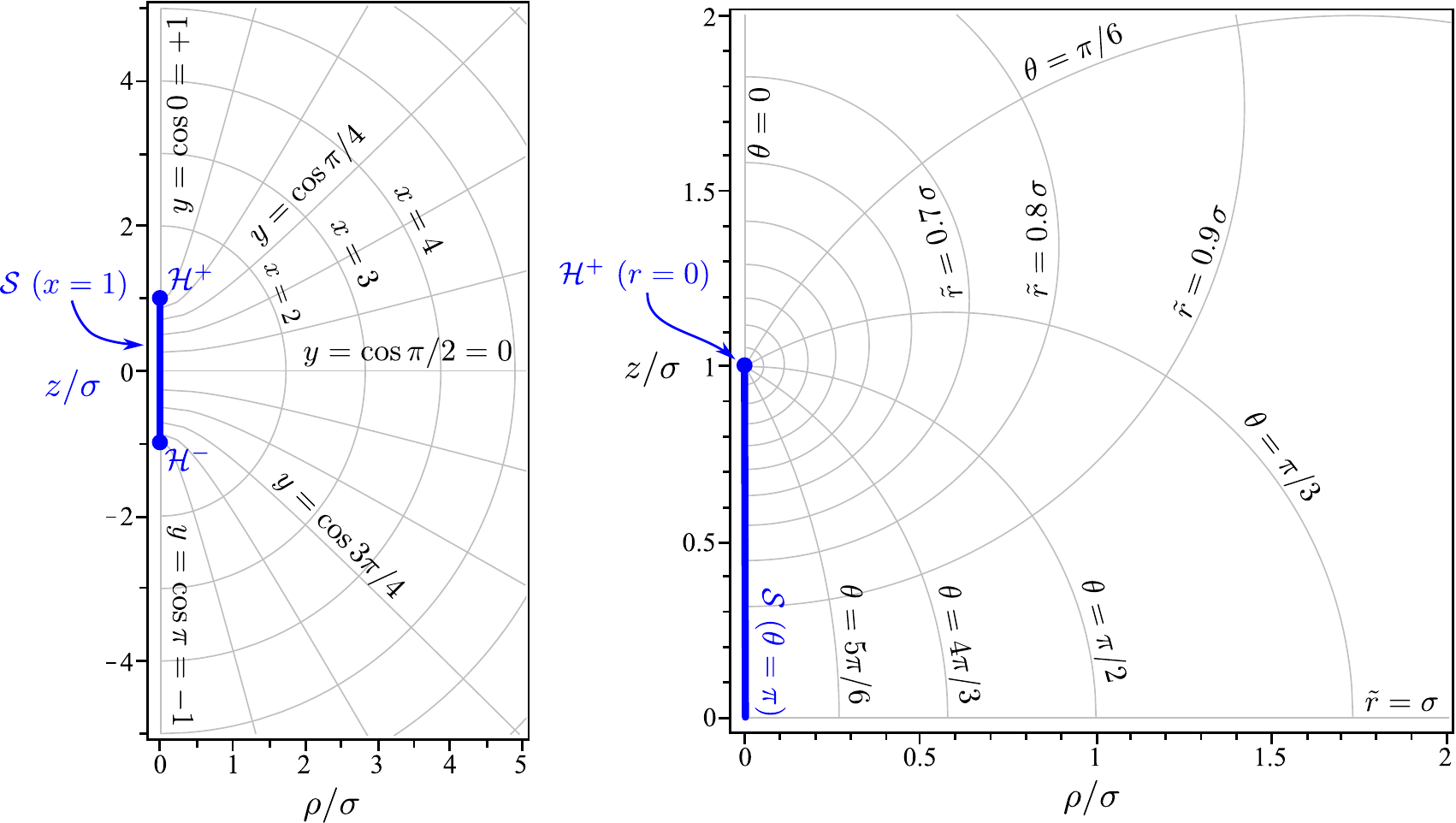}
\caption{Coordinate traces of the prolate spheriodal $(x,y)$ (\emph{left}) and bipolar $(\tilde{r},\theta)$ (\emph{right}) coordinates in the cylindrical $(\rho,z)$ plane.  Note we have taken positive square roots in (\ref{eq:bipolar}), which means that the bipolar coordinates only cover the upper-half of the $(\rho,z)$ plane.} \label{fig:coords}}

\subsection{Far field region}

In the limit $x \rightarrow \infty$, the line element reduces to
\begin{multline}
	ds^{2} = -\left[1-\frac{4}{px} + \mathcal{O}\left( \frac{1}{x^{2}} \right ) \right] \left[ d\tilde{t}^{2} + \frac{16\sigma q(1-y^{2})}{p^{2}x} d\tilde{t} \,d\phi\right]  \\ + \sigma^{2} \left[1+ \frac{4}{px} + \mathcal{O}\left( \frac{1}{x^{2}} \right ) \right] \left[ dx^{2} + \frac{x^{2}dy^{2}}{1-y^{2}}  + x^{2}(1-y^{2})d\phi^{2} \right].
\end{multline}
Making the following identifications:
\begin{equation}
	R = \sigma x, \quad y = \cos\vartheta, \quad M = \frac{2\sigma}{Gp}, \quad J = q GM^{2} = \frac{4\sigma^{2}q}{Gp^{2}},
\end{equation}
puts this in the form
\begin{multline}
	ds^{2} \approx -\left( 1-\frac{2GM}{R}  \right) \left( d\tilde{t}^{2} + \frac{4GJ\sin^{2}\vartheta}{R} d\tilde{t} \,d\phi\right) + \\ \left( 1+ \frac{2GM}{R} \right) \left[ dR^{2} + 
	R^{2}(d\vartheta^{2} +\sin^{2}\vartheta\,d\phi^{2}) \right].
\end{multline}
This matches what we would expect for an aysmptotically flat metric about a matter distribution of ADM mass $M$ and angular momentum $J$ (we use units in which $\hbar = c = 1$).  From the above identification we can infer that to ensure that $R$ and $M$ are positive, we should enforce
\begin{equation}
	\sigma >0, \quad 0 \le p \le 1.
\end{equation}
Furthermore, we see that the sign of $q$ determines whether the angular momentum is pointing in the positive or negative $z$-direction.  Without loss of generality, we can orient the $z$-axis such that the body has positive rotation and hence $q$ is non-negative.  Then, we put
\begin{equation}
	q = \sqrt{1-p^{2}}
\end{equation}
throughout the following discussion.

\subsection{Killing vectors and stationary observers}

The metric (\ref{eq:line element}) has two obvious Killing vectors:
\begin{equation}
	\tilde{t}^\alpha\di_{\alpha} = \di_{\tilde{t}}, \quad \phi^\alpha \di_{\alpha} = \di_{\phi}.
\end{equation}
We have the following:
\begin{gather}\nonumber
	\tilde{t}^\alpha \tilde{t}_\alpha = -f(x,y), \quad \phi^\alpha \phi_\alpha = -f(x,y)\omega^{2}(x,y) + \frac{\sigma^{2}(x^{2}-1)(1-y^{2})}{f(x,y)}, \\ \tilde{t}^{\alpha}\phi_{\alpha} = f(x,y)\omega(x,y).
\end{gather}
These Killing vectors can be used to define the 4-velocity of stationary observers:
\begin{equation}
	u^{\alpha} = \kappa^{-1}(\tilde{t}^{\alpha}+\Omega \phi^{\alpha}), \quad u^{\alpha} u_{\alpha} = -1,
\end{equation}
Here, $\Omega$ is a constant representing the observer's angular velocity and $\kappa$ is a normalization factor.  The local geometry seen by such an observer will be time-independent, hence the name ``stationary''.  In the Kerr and Schwarzschild geometries, such observers cannot exist in the trapped regions bounded by event horizons; i.e., there is no $\Omega \in \mathbb{R}$ such that $u^{\alpha}$ is timelike inside an event horizon.  
\begin{figure}
\begin{minipage}{0.48\textwidth}\vspace{-0mm}
\includegraphics[width=0.92\textwidth]{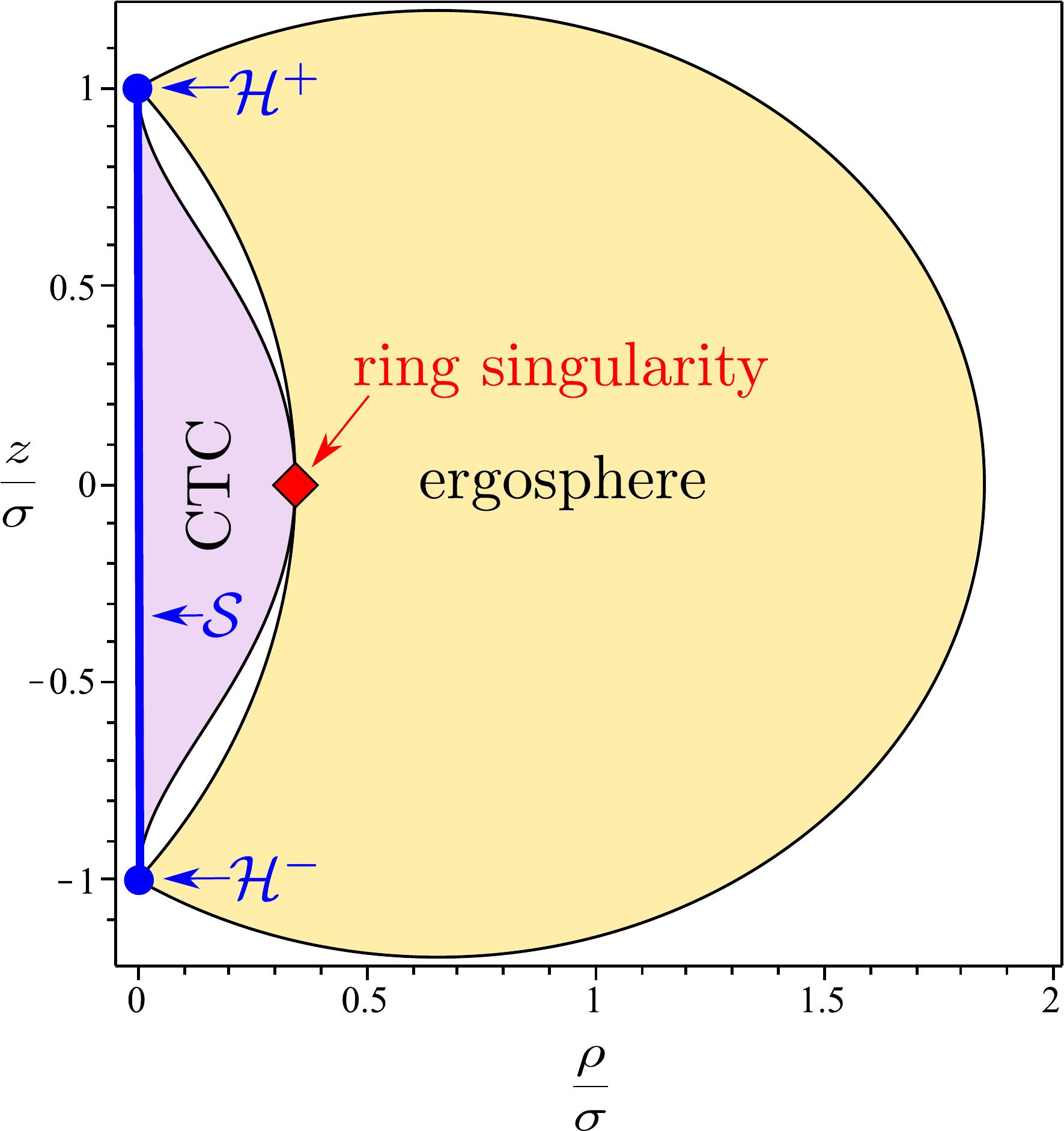}\vspace{4mm}
\caption{The ergosphere (tan) and the region with closed timelike curves (CTCs) (purple) when $p = 0.8$.  The boundaries of the ergoshere are lines $\theta =$ constant.}\label{fig:horizons}
\end{minipage} \quad
\begin{minipage}{0.48\textwidth}
\includegraphics[width=\textwidth]{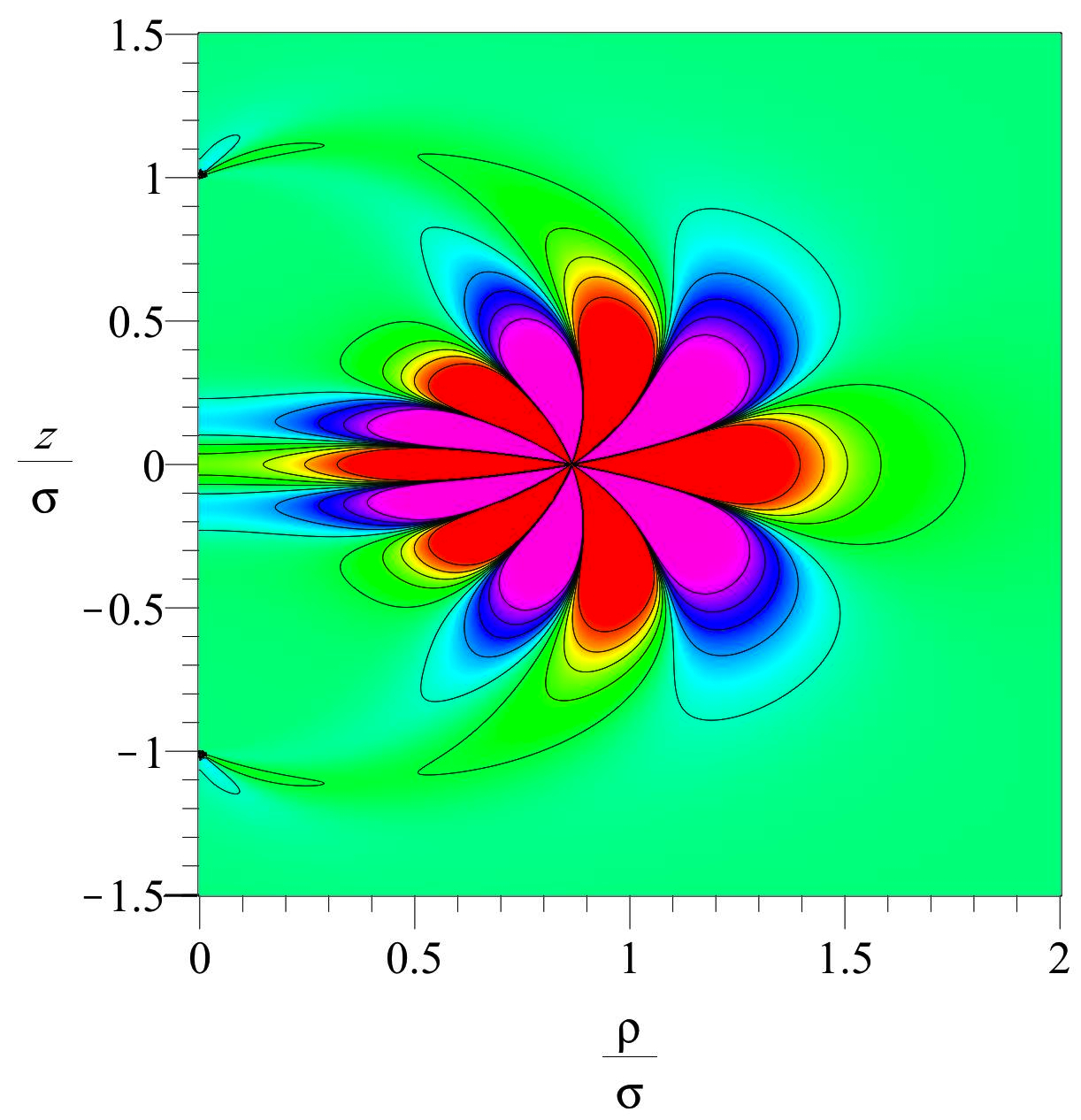}
\caption{Contour plot of the Kretchmann scalar $\tanh(\sigma^{-4} R_{\alpha\beta\gamma\delta}R^{\alpha\beta\gamma\delta})$ in the $(\rho,z)$ plane for $p=1/3$.  The centre of the flower-shaped region is the ring singularity.}\label{fig:Kret}
\end{minipage}
\end{figure}

Generally, stationary observers will exist if we can choose $\Omega \in \mathbb{R}$ such that $\kappa^{2} > 0$, where
\begin{equation}
	\kappa^{2} = -\tilde{t}^{\alpha}\tilde{t}_{\alpha} -2\Omega \tilde{t}^{\alpha}\phi_{\alpha} - \Omega^{2} \phi^{\alpha}\phi_{\alpha}.
\end{equation}
In the TS2 geometry,  this becomes
\begin{equation}
	\kappa^{2} = f(x,y)[1-\omega(x,y)\Omega]^{2}-\frac{\sigma^{2}(x^{2}-1)(1-y^{2})\Omega^{2}}{f(x,y)}. 
\end{equation}
The righthand side is a quadratic equation in $\Omega$, and it follows that if this equation has two distinct real roots there will be values of $\Omega$ for which $\kappa^{2} > 0$; i.e., values for which stationary observers exist.  Hence, the condition for the existence of stationary observers is just that the discriminant of the quadratic be positive, which reduces to
\begin{equation}
	\sigma^{2}(x^{2}-1)(1-y^{2}) = \rho^{2} > 0.
\end{equation}
In other words, there is no trapped region anywhere in the portion of the TS2 manifold covered by the $(x,y)$ coordinates, \emph{excluding} $\mathcal{S}$ and $\mathcal{H}^\pm$.  

However, there is an ergosphere in the TS2 geometry where one is obliged to have $\Omega > 0$; i.e. stationary observers are forced to rotate. This ergosphere is the portion of the manifold where $\tilde{t}^\alpha$ is spacelike.  It's boundary is the hypersurface where $\tilde{t}^\alpha$ is null; i.e., the locus of points where
\begin{equation}
	f(x,y) = A(x,y) = 0.
\end{equation}
One can show that the boundary of the ergosphere is two $\theta=$ constant hypersurfaces in the bipolar coordinates (\ref{eq:bipolar}). There is also a portion of the manifold where the angular Killing vector $\phi^\alpha$ becomes timelike.  Since $\phi$ is a periodic coordinate, this region involves closed timelike curves (CTCs) and is hence acausal.  The shape of these two regions in illustrated in Fig.\ \ref{fig:horizons}.

\subsection{The naked ring singularity}

From examination of the Kretschmann scalar $R_{\alpha\beta\gamma\delta} R^{\alpha\beta\gamma\delta}$ (which is plotted in Fig.\ \ref{fig:Kret}), we see that there is a curvature singularity in the manifold when:\footnote{It is not difficult to confirm that there are no other curvature singularities for $x \in [1,\infty)$ and $y \in [-1,1]$.}
\begin{equation}
	p^{2}x^{4}+2px^{3} - 2px - 1 = 0.
\end{equation}
One can analytically demonstrate that there is always only one root of this polynomial $x_{\ring} = x_{\ring}(p)$ with $x_{\ring} \in [1,\infty)$ for $p \in [0,1]$.  The position on this singularity is coincident with the intersection of the inner boundary of the ergosphere and the $x$-axis.  It is possible to write a closed form expression for $x_{\ring}$, but it is complicated and not illuminating, but one can derive the following limits:
\begin{align}
	\nonumber x_{\ring}(p) & = 1 + \frac{1-p}{4} + \mathcal{O}[(1-p)^{2}], \\ x_{\ring}(p) & = \frac{a}{p^{1/3}} + \mathcal{O}(p^{1/3}),
\end{align}
where $a \approx 0.79$.  So for $p$ approaching 1, the ring singularity is on the $\rho = 0$ line and for $p$ approaching 0 it is at infinity.  We can construct series expansions of various metric coefficients about the ring singularity.  Defining $\tilde{x} = x-x_{\ring}$, we find
\begin{subequations}
\begin{align}
	A & = a_{1,0} \tilde{x} + \mathcal{O}(\tilde{x}^{2},y^{2}), \\
	B & = b_{2,0} \tilde{x}^{2}  + b_{0,2} y^{2} + \mathcal{O}(\tilde{x}^{3},\tilde{x} y^{2}), \\
	C & = c_{1,0} \tilde{x} + \mathcal{O}(\tilde{x}^{2},y^{2}),
\end{align}
\end{subequations} 
where $a_{i,j}$, etc are complicated nonzero functions of $p$.  Notice that $B$ approaches zero faster than $A$ and $C$ near the singularity.  From this it follows that $f$ diverges while $\omega$ remains finite on the singularity.  Since $x_{\ring} > 1$, we conclude that stationary observers can exist arbitrarily close to the ring singularity and it must be a naked.  It is also interesting to note that as stationary observers approach the ring singularity from outside the CTC region, their angular velocity must approach a fixed value:
\begin{equation}
	\Omega \,\, \rightarrow \,\, \Omega_{\ring} = \Omega_{\ring}(p) = \frac{1}{\omega(x_{\ring}(p),0)}.
\end{equation}
This is analogous to the phenomena of stationary observers forced to co-rotating with the black hole as they approach the Killing horizon in the Kerr geometry, although here the observers approach the singularity directly.  However, if they approach the ring singularity from inside the CTC region, their angular velocity is unconstrained.

\subsection{Distribution of mass and angular momentum}\label{sec:M and J}

Extending the results of \cite{Kodama:2003ch}, we now use the Komar formulae to determine where the mass and angular momentum of the TS2 geometry resides.  Generally speaking, the mass or angular momentum contained within a closed 2-surface $\di\Sigma$ is given by
\begin{subequations}\label{eq:Komar}
\begin{align}
	M_{\di\Sigma} & = -\frac{1}{16\pi G} \int_{\di\Sigma} \epsilon_{\alpha\beta\gamma\delta} \nabla^{\gamma}\tilde{t}^{\delta} \, dx^{\alpha} \wedge dx^{\beta}, \label{eq:Komar M} \\
	J_{\di\Sigma} & = +\frac{1}{32\pi{G}} \int_{\di\Sigma} \epsilon_{\alpha\beta\gamma\delta} \nabla^{\gamma}\phi^{\delta} \, dx^{\alpha} \wedge dx^{\beta}, \label{eq:Komar J}
\end{align}
\end{subequations}
where $\tilde{t}^{\alpha}$ and $\phi^{\alpha}$ are the time and rotational Killing vectors, respectively.  If we take $\di\Sigma$ to be a surface of constant $x = x_{0}>1$, the Komar integrals reproduce the ADM mass and angular momentum:
\begin{equation}
	M_{x=x_{0}} = M, \quad J_{x=x_{0}} = J.
\end{equation}
This means that all of the mass and angular momentum of the spacetime is contained within $\mathcal{H}^{\pm}$ and $\mathcal{S}$; i.e., the ring singularity itself carries no mass or angular momentum.  We can actually explicitly calculate the mass and angular momentum contained within $\mathcal{H}^{\pm}$ by first transforming (\ref{eq:line element}) to the bipolar coordinates (\ref{eq:bipolar}).  In these coordinates, $\mathcal{H}^{\pm}$ corresponds to the $\tilde{r} = 0$ hypersurface, and we obtain
\begin{equation}
	M_{\pm} = 0, \quad J_{\pm} = \frac{p}{4(1-p)} J = \frac{\sigma^{2}}{Gp} \sqrt{\frac{1+p}{1-p}}.
\end{equation}
Interestingly, the $\mathcal{H}^{\pm}$ surfaces contain zero mass but non-zero angular momentum.  Finally, we define\footnote{Note that calculating $M_{\mathcal{S}}$ and $J_{\mathcal{S}}$ from (\ref{eq:Komar}) directly is problematic because the $x = 1$ hypersurface overlaps with some portion of $\mathcal{H}^{\pm}$.} 
\begin{equation}
	M_{\mathcal{S}} = M - 2 M_{\pm} = M, \quad J_{\mathcal{S}} = J - 2 J_{\pm} =  \left( 1 - \frac{p}{2} \right) J.
\end{equation}
That is, all the mass but only a fraction of the total angular momentum is carried within $\mathcal{S}$.

\subsection{Killing horizons and the near horizon metric}

To further investigate the geometry near $\mathcal{H}^\pm$, we again change coordinates from $(x,y)$ to $(\tilde{r},\theta)$ in the line element (\ref{eq:line element}).  For concreteness, we take positive square roots in (\ref{eq:bipolar}) which means that $\tilde{r}=0$ corresponds to $\mathcal{H}^{+}$.  The immediate neighbourhood of $\mathcal{H}^+$ is defined by $\tilde{r} \ll \sigma$, and in this regime the line element reduces to:
\begin{subequations}\label{eq:NHTS2}
\begin{gather}
	ds^{2} = \tilde\Gamma(\theta) \left( -\frac{\tilde{r}^{2}}{4r_{0}^{2}} d\tilde{t}^{2} +  \frac{r_{0}^{2}}{\tilde{r}^{2}} d\tilde{r}^{2} + r_{0}^{2} d\theta^{2} \right) + \frac{\sin^{2}\theta}{\tilde\Gamma(\theta)}  \left( \frac{q \tilde r}{2 p r_{0}} d\tilde{t} -r_{0}d\phi \right)^{2}, \\
	\tilde\Gamma(\theta) \equiv \frac{\cos^{2}\theta + 2(2p^{2}-1)\cos\theta+1}{4p^{2}}, \\
	r_{0}^{2} \equiv \frac{2\sigma^{2}(p+1)}{p^{2}} = \frac{G^{2}M^{2} + \sqrt{G^{4}M^{4}-G^{2}J^{2}}}{2} = \frac{2GJ_{\pm}\sqrt{1-p^{2}}}{p}.\label{eq:TS2 r0 def}
\end{gather}
\end{subequations}
We call (\ref{eq:NHTS2}) the ``near horizon Tomimatsu-Sato'' (NHTS) geometry.  If we change coordinates according to
\begin{equation}
	\tilde{r} = \sqrt{2} \bar{r}, \quad \tilde{t} = \sqrt{2} \bar{t},
\end{equation}
we see that the NHTS geometry is a special case of the following spacetime:
\begin{subequations}\label{eq:NHES}
\begin{gather}
	ds^{2} = \Gamma(\theta) \left( -\frac{\bar{r}^{2}}{r_{0}^{2}} d\bar{t}^{2} +  \frac{r_{0}^{2}}{\bar{r}^{2}} dr^{2} + r_{0}^{2} d\theta^{2} \right) + \frac{\sin^{2}\theta}{\Gamma(\theta)}  \left( \frac{\gamma \bar{r}}{r_{0}} d\bar{t} -r_{0}d\phi \right)^{2}, \label{eq:NHES metric} \\
	\Gamma(\theta) = \tfrac{1}{2} \alpha(\cos^{2}\theta + 1) + \beta \cos\theta, \quad \gamma^{2} = \alpha^{2} - \beta^{2}. \label{eq:Gamma ansatz}
\end{gather}
\end{subequations}
We call this the generic ``near-horizon extremal spinning'' (NHES) geometry.  It is easily confirmed that this metric has $R_{\alpha\beta} = 0$ and is free from curvature singularities for $\gamma \ne 0$.  As demonstrated in Appendix \ref{app:NHES}, the above line element (\ref{eq:NHES}) is isometric to the most general vacuum solution representing the near horizon (non-toriodal) geometry of an axisymmetric extremal horizon in 4-dimensional general relativity \cite{Kunduri:2008rs}.  For various choices of the parameters $(\alpha,\beta,r_{0})$, this metric describes the near-horizon Tomimatsu-Sato, extremal-Kerr \cite{Bardeen:1999px}, or the extremal Kerr-bolt \cite{Ghezelbash:2009gy} solutions (the exact parameters corresponding to each of these parent geometries are given in Table \ref{tab:NHES}).  
\TABLE[t]{
\newcolumntype{Q}{>{$\displaystyle}m{45mm}<{$}}
\newcolumntype{P}{>{$}m{38mm}<{$}}
\newcolumntype{Z}{>{$\displaystyle}m{23mm}<{$}}
\begin{tabular}{PZZZZ}
\hline \text{Parent solution} & \alpha
& \beta & \gamma & r_{0}^{2}  \\
\hline \hline%
 
\text{TS2} & \frac{1}{2p^{2}} & 1-\frac{1}{2p^{2}} & \left( \frac{1}{p^{2}}-1 \right)^{1/2} & 2\gamma GJ_{\Delta} \\

\text{Extremal Kerr \cite{Bardeen:1999px}} & 1 & 0 & 1 & 2GJ \\

\text{Extremal Kerr-bolt \cite{Ghezelbash:2009gy}} & 1 & \frac{N}{a} & \left( 1-\frac{N^{2}}{a^{2}}\right)^{1/2} & 2a^{2} \\

\hline
\end{tabular}
\caption{NHES metric parameters for the near-horizon limit of the TS2, extremal Kerr, and extremal Kerr-bolt geometries.  Here, $N$ is the NUT charge of the Kerr-bolt geometry and $a = J/M$.  Notice that all of the solutions have $\gamma^{2} = \alpha^{2}-\beta^{2}$, which means they are all Ricci flat.} 
\label{tab:NHES}
}

It is obvious that the metric (\ref{eq:NHES}) has a Killing horizon at $\bar{r} = 0$ that we denote by $\Delta$ (and which corresponds to $\mathcal{H}^{\pm}$ in the full TS2 geometry) generated by the null Killing vector $\ell^{\alpha} \di_{\alpha} = \di_{\bar{t}}$.  The surface gravity of the horizon $\kappa$ can be deduced from the relation $\ell^{\alpha} \nabla_{\alpha} \ell^{\beta} = \kappa \ell^{\beta}$, which gives $\kappa = 0$; that is, the horizon is extremal.  We can extend the solution across the horizon by changing to the global coordinates: 
\begin{subequations}
\begin{align}
	\bar{r} & = r_{0} ( \sqrt{1+r^{2}} \cos t + r), \\ \bar{t} & = \frac{r_{0}^{2}\sqrt{1+r^{2}}\sin t}{\bar{r}}, \\ \phi & = -\gamma\left[ 
	\varphi + \ln \left( \frac{\cos t + r \sin t}{1 + \sqrt{1+r^{2}}\sin t} \right) \right].
\end{align}
\end{subequations}
In terms of these \emph{dimensionless} coordinates $(t,r,\theta,\varphi)$, the NHES line element is
\begin{equation}
	ds^{2} = r_{0}^{2} \Gamma(\theta) \left[  -(1+r^{2}) dt^{2} + \frac{dr^{2}}{1+r^{2}} + d\theta^{2} + \frac{\gamma^{2}\sin^{2}\theta}{\Gamma^{2}(\theta)} \left( d\varphi + r\,dt \right)^{2} \right].
\end{equation}
Notice that the period of the new angular coordinate is \emph{not} $2\pi$:
\begin{equation}
	\phi \sim \phi + 2\pi, \quad \varphi \sim \varphi + \frac{2\pi}{\gamma}.
\end{equation}
Just as for the NHEK and the NHTB solutions, the NHES metric (\ref{eq:NHES}) has $SL(2,\mathbb{R})\times U(1)$ symmetry.  The $SL(2,\mathbb{R})$ isometry is generated by the three Killing vectors
\begin{subequations}
\begin{align}
	K_{1}^{\alpha}\di_{\alpha} & = +2 r (1+r^{2})^{-1/2}\sin t \, \di_{t} - 2 (1+r^{2})^{1/2} \cos t \, \di_{r} + 2 (1+r^{2})^{-1/2} \sin t \, \di_{\varphi}, \\  K_{2}^{\alpha}\di_{\alpha} & =  -2 r (1+r^{2})^{-1/2}\cos t \, \di_{t} - 2 (1+r^{2})^{1/2} \sin t \, \di_{r} - 2 (1+r^{2})^{-1/2} \cos t \, \di_{\varphi}, \\ K_{0}^{\alpha}\di_{\alpha} & = +2\,\di_{t},
\end{align}
\end{subequations}
which satisfy the algebra:
\begin{equation}
	[K_{1},K_{2}] = 2K_{0}, \quad [K_{0},K_{2}] = 2K_{1}, \quad [K_{0},K_{1}] = -2K_{2}.
\end{equation}
The $U(1)$ rotational isometry is generated by the Killing vector $\bar{K}_{0}^{\alpha}\di_{\alpha} = -\di_{\varphi}$.

The metric and area on the horizon are easy to calculate:
\begin{equation}
	ds_{\Delta}^{2} = r_{0}^{2 }\left[ \Gamma(\theta) d\theta^{2} + \frac{\sin^{2}\theta}{\Gamma(\theta)}d\phi^{2} \right], \quad A_{\Delta} = 4\pi r_{0}^{2},
\end{equation}
which implies that the \emph{total} horizon area in TS2 is
\begin{equation}
	A_{\text{TS2}} = 2 \times A_{\Delta} = 4\pi [ G^{2}M^{2} + \sqrt{G^{4}M^{4}-G^{2}J^{2}} ].
\end{equation}
We see that the combined area of the TS2 horizons is precisely half that of a Kerr black hole with the same mass and angular momentum, consistent with the results of Ref.\ \cite{Kodama:2003ch}.

We now determine the mass and angular momentum of $\Delta$.  A well-defined algorithm for calculating these entirely in terms of quantities defined on $\Delta$ is the isolated horizon formalism \cite{Ashtekar:2000sz}.  Since we are dealing with a vacuum spacetime, the mass and angular momentum are defined as
\begin{subequations}
\begin{align}\label{eq:IH J}
	J_\Delta &  = +\frac{1}{32\pi{G}} \int_{\Delta} \epsilon_{\alpha\beta\gamma\delta} \nabla^{\gamma}\phi^{\delta} \, dx^{\alpha} \wedge dx^{\beta} = \frac{r_{0}^{2}}{2\gamma G}, \\
	M_{\Delta}^{2} & = \frac{r_{0}^{4} + 2G^{2}J_{\Delta}^{2} }{4G^{2}r^{2}_{0}} = \frac{(1+\gamma^{2})J_{\Delta}}{2\gamma G},
\end{align}
\end{subequations}
where $\phi^{\alpha}$ is the Killing vector generating the rotational isometry, as before.  Other than the fact that the above quantities are to be calculated using the NHES metric, we note that (\ref{eq:IH J}) is identical to the Komar angular momentum (\ref{eq:Komar J}) we defined above for the $\mathcal{H}^{\pm}$ surfaces in the TS2 manifold.  Hence, it is not surprising that $J_{\Delta} = J_{\pm}$ when (\ref{eq:TS2 r0 def}) is enforced.  However, we note that $M_{\Delta} \ne M_{\pm}$.  Again, this is not surprising: $M_{\pm}$ can be thought of as the charge of the horizon conjugate to the generator of asymptotic time translations; i.e., the ADM mass.  However, as pointed out in \cite{Ashtekar:2000sz}, there is no guarantee that the isolated horizon mass agree with the ADM mass since quantities defined on the horizon do not necessarily have any knowledge of the asymptotically flat portion of the spacetime.  For $\gamma = 1$, we recover the Kerr-extremality condition $J_{\Delta} = GM^{2}_{\Delta}$.

Finally, we note that for generic choices of $\alpha$ and $\beta$, the horizons $\Delta$ have conical singularities on the north ($\theta = 0$) and south ($\theta = \pi$) poles.  The conical deficit (or excess) about each pole is:
\begin{equation}
	\delta_{\mathrm{N},\mathrm{S}} = 2\pi \left[ 1-(\alpha \pm \beta)^{-1} \right].
\end{equation}
Referring to Table \ref{tab:NHES}, we see that the NHEK metric has no singularities, the NHTS metric has a defect at the south pole, while the NHKB metric has defects at both poles.  Also note that if $\gamma \ne 1$, the NHES geometry necessarily involves conical singularities.

\subsection{Nature of the $\mathcal{S}$ hypersurface}

We now attempt to further understand the nature of the $\mathcal{S}$ hypersurface.  First, we note that the Killing vectors $\tilde{t}^{\alpha}$ and $\phi^{\alpha}$ are both timelike in the neighbourhood of $\mathcal{S}$ and it is easy to confirm
\begin{equation}
	\lim_{x\rightarrow 1} [ (\tilde{t}^{\alpha }\tilde{t}_{\alpha})( \phi^{\beta}\phi_{\beta}) - (\tilde{t}^{\alpha}\phi_{\alpha})^{2}] = 0;
\end{equation}  
i.e., the Killing vectors are parallel on $\mathcal{S}$.  In fact, it is not hard to find a linear combination that vanishes in the $x \rightarrow 1$ limit:
\begin{equation}
	\lim_{x \rightarrow 1} (\tilde{t}_{\alpha} + \Omega_{\mathcal{S}} \phi_{\alpha}) = 0, \quad  \Omega_{\mathcal{S}} = \frac{p}{4\sigma}\sqrt{\frac{1-p}{1+p}}.
\end{equation}
Next, following \cite{Kodama:2003ch} let us introduce new coordinates
\begin{equation} 
	\tau = \frac{\Omega_{\mathcal{S}}\tilde{t} - \phi}{\Omega_{\mathcal{S}}}, \quad x = \cosh\eta.
\end{equation}
In these coordinates, $\mathcal{S}$ corresponds to $\eta = 0$.  Expanding the TS2 line element in the $\eta \ll 1$ regime yields
\begin{multline}
	ds^{2} = - \frac{(1-y^{2})^{2}(1-p)}{g(y)} d\tau^{2} + \frac{(1-p^{2})(1+p)g(y)\sigma^{2}}{p^{4}(1-y^{2})}\left( 
d\eta^{2} + \frac{dy^{2}}{1-y^{2}}\right) \\ + \frac{\eta^{2}g(y) \sigma^{2}}{(1-p)(1-y^{2})} d\phi^{2} +  \frac{4\sigma\eta^{2}[g(y)-8y^{2}(p+1)] }{g(y)\sqrt{1-p^{2}}} d\tau\,d\phi, 
\end{multline}
where
\begin{equation}
	g(y) \equiv (1-p)y^{4} + 2(p+3)y^{2} + 1 -p >0.
\end{equation}
Examination of the $(t,y)=$ constant sections reveals there is a conical singularity about $\mathcal{S}$, and the deficit angle is
\begin{equation}
	\delta = \frac{2\pi}{1-p^{2}}.
\end{equation}
This is the same deficit angle as we found above in each of the horizons $\mathcal{H}^{\pm}$.  The metric on $\mathcal{S}$ itself is seen to be
\begin{equation}
	ds_{\mathcal{S}}^{2} = - \frac{(1-y^{2})^{2}(1-p)}{g(y)} d\tau^{2} + \frac{(1-p^{2})(1+p)g(y)\sigma^{2}}{p^{4}(1-y^{2})^{2}} dy^{2}. 
\end{equation}
We recognize a Lorentzian $1+1$ spacetime, which demonstrates $\mathcal{S}$ is a line-like object.  To summarize the results of this section and \S\ref{sec:M and J} above, we have seen that $\mathcal{S}$ is a one-dimensional object carrying non-zero mass and angular momentum.  It is coincident with a conical singularity and is surrounded by closed timelike curves.

\section{CFT description of the near horizon geometry}\label{sec:CFT}

In this section, we seek a CFT dual to the generic near horizon geometry (\ref{eq:NHES}) defined above.  We will find that the existence of such a CFT is plausible, and that the basic parameters of the theory are given most concisely in terms of near horizon quantities.

\subsection{Central charge}

We first find the central charge of the candidate dual CFT  by analyzing the asymptotic symmetries of metric fluctuations using the method of Ref.\  \cite{Guica:2008mu} based on the covariant formalism of Refs.\ \cite{Barnich:2001jy,Barnich:2007bf}.\footnote{The CFT description of very similar near horizon geometries arising from extremal black holes of varying matter content has been been obtained in Refs.\ \cite{Hartman:2008pb,Compere:2009dp}.}  We assume the NHES metric is perturbed according to $g_{\alpha\beta} \rightarrow g_{\alpha\beta} + h_{\alpha\beta}$ and assume the same boundary conditions on $h_{\alpha\beta}$ as in Ref.\ \cite{Guica:2008mu}.  The most general diffeomorphism consistent with the assumed boundary conditions has the following form as $r \rightarrow \infty$
\begin{subequations}
\begin{align}
	\xi^{\alpha} & = \zeta^{\alpha} + \chi^{\alpha}, \\ \zeta^{\alpha} \di_{\alpha} & = [\epsilon(\varphi)+\mathcal{O}(r^{-2})] \di_{\varphi} - [r \epsilon'(\varphi)+\mathcal{O}(r^{0})] \di_{r}, \\ \chi^{\alpha} \di_{\alpha} & = [C+\mathcal{O}(r^{-3})]\di_{\tau},
\end{align}
\end{subequations}
where $\epsilon(\varphi)$ and $C$ are an arbitrary function and constant, respectively.  If we choose
\begin{equation}
	\epsilon(\varphi) = \frac{1}{\gamma} e^{i\gamma m\varphi}, \quad \epsilon(\varphi) = \epsilon(\varphi + 2\pi/\gamma), \quad  m = 0, \pm 1, \pm 2, \ldots 
\end{equation}
and drop subleading terms, we see that the generators 
\begin{equation}
	\zeta_{m} = \gamma^{-1} e^{i\gamma m \varphi} [ \di_{\varphi} - i\gamma m r \di_{r}],
\end{equation}	
satisfy the Virasoro algebra:
\begin{equation}\label{eq:Virasoro}
	i[\zeta_{m},\zeta_{n}] = (m-n)\zeta_{m+n}.
\end{equation}
Defining
\begin{multline}
	k_{\zeta}[h,g] = -\tfrac{1}{4} \epsilon_{\alpha\beta\mu\nu} [ \zeta^{\nu} \nabla^{\mu} h - \zeta^{\nu}\nabla_{\sigma} h^{\mu\sigma} + \tfrac{1}{2} h \nabla^{\nu}\zeta^{\mu} \\ - h^{\nu\sigma}\nabla_{\sigma} \zeta^{\mu} + \tfrac{1}{2} h_{\sigma}{}^{\nu}( \nabla^{\mu}\zeta^{\sigma} +  \nabla^{\sigma}\zeta^{\mu}  ) ] dx^{\alpha} \wedge dx^{\beta},
\end{multline}
the charges $Q_{\zeta_{m}}$ associated with the diffeomorphisms $\zeta_{m}$ satisfy the algebra
\begin{equation}
	\{ Q_{\zeta_{m}}, Q_{\zeta_{n}} \} = -i(m-n)Q_{\zeta_{m}+\zeta_{n}} + \frac{1}{8\pi G} \int_{\di\Sigma} k_{\zeta_{m}}[\pounds_{\zeta_{n}}g,g],
\end{equation}
where $\{,\}$ is the Dirac bracket.  Here, $\di\Sigma$ is a $(t,r) =$ constant hypersurface in the limit $r \rightarrow \infty$.  Evaluating the integral in the NHES background yields
\begin{equation}
	i \{ Q_{\zeta_{m}}, Q_{\zeta_{n}} \} = (m-n)Q_{\zeta_{m}+\zeta_{n}} + \frac{c}{12} m \left( m^{2}+\frac{2}{\gamma^{2}} \right) \delta_{m+n,0},
\end{equation}
where the central charge $c$ is given by
\begin{equation}\label{eq:central charge 1}
	c = \frac{6\gamma r_{0}^{2}}{G} = 12 \gamma^{2} J_{\Delta}.
\end{equation}
Passing over to the quantum theory, we make the switch $\{,\} \rightarrow -i[,]$ and define $L_{n}$ as the quantum versions of the $Q_{\zeta_{n}}$:
\begin{equation}
	L_{n} = Q_{\zeta_{n}} + \frac{c}{12} \left( \frac{1}{\gamma^{2}} - \frac{1}{2} \right) \delta_{n,0},
\end{equation}
which yields the familiar expression
\begin{equation}
	[L_{m},L_{n}] = (m-n)L_{m+n} + \frac{c}{12} m (m^{2}-1)\delta_{m+n,0}.
\end{equation}

Our formula for the central charge (\ref{eq:central charge 1}) reproduces the NHEK \cite{Guica:2008mu} and NHKB \cite{Ghezelbash:2009gy} results with appropriate identifications listed in Table \ref{tab:NHES}.  Indeed, it looks very much like the original extremal-Kerr result $c = 12J$, apart from the $\gamma$ factor and the appearance of the horizon angular momentum $J_{\Delta}$ instead of the ADM value $J$.  Of course, in Kerr such a distinction is irrelevant because $J = J_{\Delta}$, but in the TS2 spacetime this is not the case.  This suggests that any CFT description that we deduce for the TS2 geometry is really a CFT description of one of the horizons.

\subsection{Temperature}

We now turn our attention to the temperature of the dual CFT.  A particularly efficient way of deriving the left and right moving temperatures involves analysis of the massless scalar wave equation $\nabla_{\alpha} \nabla^{\alpha} \Phi = 0$ in the near horizon region (\ref{eq:NHES}) \cite{Castro:2010fd,Chen:2010fr}.\footnote{We should make it clear that the following calculation is performed entirely in the near-horizon region, as in \cite{Castro:2010fd}, not throughout the entire spacetime.  For the full TS2 parent geometry, the Cauchy problem for the evolution of a scalar field is ill-defined due to the presence of the naked singularity and the CTC region.}  Let us make a separation of variables \emph{ansatz}:
\begin{equation}
	\Phi(\bar{t},\bar{r},\theta,\phi) = e^{-i\omega \bar{t} + im\phi} S(\theta) R(\bar{r}).
\end{equation}
We find the wave equation reduces to:
\begin{subequations}
\begin{align}
	-K S(\theta) & = \left[\di_{\theta}^{2} + \frac{\cos\theta}{\sin\theta} \, \di_{\theta} + m^{2}\gamma^{2} - \frac{m^{2} \Gamma(\theta)}{ \sin^{2}\theta} \right]S(\theta), \\
	K R(\bar{r}) & = \left[ \bar{r}^{2 }\di_{\bar{r}}^{2}  + 2\bar{r} \di_{\bar{r}} + \frac{r_{0}^{4}\omega^{2}}{\bar{r}^{2}}  - \frac{2\gamma m \omega r_{0}^{2}}{r} \right] R(\bar{r}),
\end{align}
\end{subequations}
where $K$ is a separation constant.  The angular equation is solvable in terms of Huen functions and yields values for $K$ once boundary conditions are enforced.  Turning our attention to the radial equation, if we change variables according to
\begin{align}\label{eq:conformal coords}
	w_{+} = -\frac{1}{2} \left( \frac{\bar{t}}{r_{0}} + \frac{r_{0}}{\bar{r}} \right),\quad
	w_{-} = \frac{1}{2}e^{\phi/\gamma}, \quad
	y = \sqrt{\frac{r_{0}}{2\bar{r}}} e^{\phi/2\gamma},
\end{align}
then the radial equation becomes (after noting $im = \di_{\phi}$ and $i\omega = -\di_{\bar{t}}$):
\begin{equation}
	K R = \left[ \tfrac{1}{4} (y^{2}\di_{y}^{2} - y \di_{y}) + y^{2} \di_{+}\di_{-} \right] R.
\end{equation}
The differential operator appearing on the right  can be expressed as
\begin{equation}
	\tfrac{1}{4} (y^{2}\di_{y}^{2} - y \di_{y}) + y^{2} \di_{+}\di_{-} = \mathfrak{H} = \bar{\mathfrak{H}},
\end{equation}
where $\mathfrak{H}$ and $\mathfrak{\bar{H}}$ are the quadratic Casimirs
\begin{equation}
	\mathfrak{H} = -H_{0}^{2} + \tfrac{1}{2} (H_{+1}H_{-1} + H_{-1}H_{+1}), \quad \bar{\mathfrak{H}} = -\bar{H}_{0}^{2} + \tfrac{1}{2} (\bar{H}_{+1}\bar{H}_{-1} + \bar{H}_{-1}\bar{H}_{+1}),
\end{equation} 
formed by the $SL(2,\mathbb{R})$ generators
\begin{subequations}
\begin{align}
	H_{+1} & = i\di_{+}, \quad H_{0} = i(w_{+}\di_{+} +\tfrac{1}{2} y \di_{y}), \quad H_{-1} = i(w_{+}^{2}\di_{+} + w_{+}y \di_{y} - y^{2}\di_{-}), \\ \bar{H}_{+1} & = i\di_{-}, \quad \bar{H}_{0} = i(w_{-}\di_{-} +\tfrac{1}{2} y \di_{y}), \quad \bar{H}_{-1} = i(w_{-}^{2}\di_{-} + w_{-}y \di_{y} - y^{2}\di_{+}),
\end{align}
\end{subequations}
respectively.  As argued in detail elsewhere \cite{Castro:2010fd}, this would seem to imply that the solutions of the wave equation are representations of $SL(2,\mathbb{R})$.  However, this is not really the case because the coordinate transformation (\ref{eq:conformal coords}) implies that the $w_{-}$ coordinate should be identified as follows:
\begin{equation}
	w_{-} \sim w_{-} e^{2\pi/\gamma}.
\end{equation}
The net effect of this identification is to break the conformal symmetry and to ensure that observers using the original $(\bar{t},\bar{r})$ coordinates (i.e., co-rotating with the horizon) will observe a thermal bath of (left-moving) radiation of temperature
\begin{equation}
	T_{L} = \frac{1}{2\pi\gamma},
\end{equation} 
from the $SL(2,\mathbb{R}) \times SL(2,\mathbb{R})$ invariant vacuum via the Unruh effect.\footnote{For the extremal situation we are considering here, the right-moving (Hawking) temperature is zero ($T_{R} = 0$) since the $w_{+}$ coordinate does not require an identification.}  This reproduces the Frolov-Thorne temperature \cite{Frolov:1989jh} of the extremal Kerr horizon when $\gamma =1$.

\subsection{Entropy}

We now briefly discuss the thermodynamics of the NHES horizon $\Delta$.  The Bekenstein-Hawking entropy of the horizon is, as usual,
\begin{equation}
	S_{\text{BH}} = \frac{A_{\Delta}}{4G} = \frac{\pi r_{0}^{2}}{G}  =  2\pi \gamma J_{\Delta}.
\end{equation}
This expression is highly analogous to to the one derived in \cite{Hartman:2008pb} for various extremal spinning black holes with magnetic and electric charges.  The form of the entropy suggests we view $J_{\Delta}$ and $\gamma$ as the thermodynamic state variables for our system.\footnote{One could equivalently view the system to be thermodynamically described by the isolated horizon quantities $(J_{\Delta},M_{\Delta})$.}  Following a similar procedure to the one in \cite{Hartman:2008pb}, a temperature can be associated with each coordinate via
\begin{equation}
	dS_{\text{BH}} = \frac{dJ_{\Delta}}{T_{J}} + \frac{d\gamma}{T_{\gamma}}.
\end{equation}
Now, since the charge $J_{\Delta}$ generated by the zero mode $\zeta_{0}$ of the Virasoro algebra (\ref{eq:Virasoro}), $T_{J}$ ought to be identified with the left-moving temperature $T_{L}$ of the dual CFT.  This yields
\begin{equation}
	T_{L} = \frac{1}{2\pi\gamma},	
\end{equation}	
consistent with the temperature derived from the broken conformal symmetry of the scalar wave equation. Putting this temperature into the Cardy formula gives the entropy of the dual CFT:
\begin{equation}
	S_{\text{CFT}} = \frac{1}{3}\pi^{2} c T_{L} = \frac{1}{3} \pi^{2} \left( \frac{6\gamma r_{0}^{2}}{G} \right) \left( \frac{1}{2\pi\gamma} \right) = \frac{\pi r_{0}^{2}}{G} = S_{\text{BH}}.
\end{equation}
Hence the macroscopic and microscopic entropies agree.

\section{Discussion}\label{sec:discussion}

In this paper, we have examined the classical properties of the $\delta = 2$ Tomimatsu-Sato geometry and then calculated the basic parameters associated with its dual CFT description using techniques from the recently proposed Kerr/CFT correspondence.  In particular, we have confirmed that the TS2 geometry involves a naked ring-like curvature singularity and a spinning rod-like conical singularity.  The former exists on the boundary of a Kerr-like ergosphere, while the latter is  surrounded by a region containing closed timelike curves.  All the ADM mass and a portion of the angular momentum of the geometry is carried by the strut.  The remaining angular momentum is carried by a pair of Killing horizons located at each end of the rod, while the ring curvature singularity contains no mass or angular momentum whatsoever.  The two Killing horizons are extremal in the sense that their surface gravity is zero, but unlike the extremal Kerr solution there exists no special relationship between the ADM mass and angular momentum at infinity.

As pointed out by Kodama and Hikida \cite{Kodama:2003ch}, there is no completely compelling evidence that the TS2 spacetime considered here cannot be the endpoint of gravitational collapse.  Indeed, there been recent work examining the prospects for observing the TS2 geometry in a realistic astrophysical context \cite{Bambi:2010hf}.  There is good motivation for determining whether or not TS2 objects exist in our universe, because their observation would provide an irrefutable counterexample to the cosmic censorship conjecture.  But it remains to be seen if the TS2 solution is stable, which is a necessary condition for such an object to be seen in nature.  The TS2 solution escapes the black hole uniqueness theorem by the existence of the naked ring singularity, so it can be viewed as an asymptotically-flat alternative geometry to Kerr.  Given $M$ and $J$, we have seen that the total horizon area of a TS2 object be half that of the matching Kerr black hole.  Applying the standard interpretation, this means that the former has half the entropy of the latter.  Thermodynamic lore then implies that the endpoint of non-equilibrium gravitational processes ought to be the Kerr black hole rather than the TS2 spacetime.  Such arguments have been used in the past to account for the Gregory-Laflamme instability of black strings.  However, as pointed out by Reall \cite{Reall:2001ag}, the existence of this type of global thermodynamic instability in a system does not imply a classical instability of the state with lower entropy.  In other words, the perturbative stability of the TS2 geometry is still an open question.  Addressing this by direct computation or by making arguments based on the local thermodynamic stability properties of the solution is an interesting avenue for future work.\footnote{An important conceptual difficulty with such a study is that the existence of the closed timelike curve region makes the Cauchy problem for perturbations ill-defined.  We thank H.\ Kodama for bringing this to our attention.}

We have demonstrated that the near horizon geometry of the TS2 model is of a form that represents the most general extremal horizon in vacuum general relativity, and  we have calculated the central charge and temperature of the CFT dual to this entire class.  In addition to providing the basic parameters for a CFT description of the TS2 horizon(s), we reproduce known results for extremal Kerr \cite{Guica:2008mu} and extremal Kerr-Bolt \cite{Ghezelbash:2009gy} solutions.  Despite the rather exotic features (such as the ring singularity) of the TS2 geometry away from the horizons, the dual CFT seems to be sensitive to only properties of the horizons themselves.  For example, the central charge is $c = 12\gamma J_{\Delta}$, where $J_{\Delta}$ is the angular momentum of the horizon only, not the entire spacetime.  In conventional spacetimes, such distinctions are irrelevant because the ADM and horizon charges usually agree.

In addition to the dependence on angular momentum, the temperature and central charge of the CFT depend on another parameter $\gamma$.  The precise physical or geometric interpretation of the quantity is unclear, but it appears to be related to the existence of singularities in the near horizon geometry; i.e., if $\gamma \ne 1$, the near horizon geometry has conical singularities and if $\gamma = 1$ we recover results for extremal Kerr.  We note that the conical singularities in the near horizon of TS2 or extremal Kerr-Bolt are very similar like the singularities of 4D static black hole solutions with multiple D6-branes in 4-dimensional $\mathcal{N}=2$ supergravity \cite{Gaiotto2006}.  So, it is natural to speculate about a stringy origin of these conical singularities, as in Refs.\ \cite{Gaiotto2006, Balasubramanian2001, Lunin2002, Maldacena2002}.  Furthermore, we know from the examples of the $\text{AdS}_3/\text{CFT}_2$ and Kerr/CFT correspondences, there are alternatives to the covariant version of the Brown-Henneaux formalism employed here to obtain the central charge.  In particular, it has been shown that the extremal Kerr central charge can be obtained by dimensionally reducing the action from four to two dimensions and the careful addition of counterterms \cite{Castro2008a, Castro2009, Castro2010}.  It would be interesting to re-do these calculations for the near-horizon extremal spinning metric considered in this work with the goal of gaining greater insight into the role of the $\gamma$ parameter.  Work in this direction is in progress. 

Finally, we can speculate what our work means for the wider Kerr/CFT programme.  We are lead to conjecture that gravity-CFT duals of the type discussed in this paper are successful at encoding information contained within a given horizon, but not sensitive to the nature of spacetime outside (including the presence of a naked singularity).  This can be a discouraging conclusion to a purist who was hoping for quantum model aware of the global properties of the spacetime, but it is an encouraging development for those who seek CFT duals of black holes embedded in the \emph{real} universe, which is obviously not asymptotically flat.  Our conjecture follows simply from the observation that the properties of the CFT we chose to examine depend only on the near horizon geometry, and the nature of that geometry is tightly constrained by the Einstein field equations under the extremality condition.  In order to confirm or refute this conclusion, it would be useful to have some understanding of other fundamental properties of the CFT; i.e., features distinct from the central charge and the temperature.  Given such knowledge, we could test if these other properties are sensitive to the detailed bulk geometry in the various spacetimes containing extremal spinning horizons (i.e.\ extremal Kerr, extremal Kerr-Bolt, and Tomimatsu-Sato $\delta=2$ solutions).  This is a promising avenue for future work.

\acknowledgments

We thank Viqar Husain and Hideo Kodama for useful conversations.  H.\ Liu thanks Wei Song for valuable discussion and also appreciates the hospitality and inspiring atmosphere of the 8th Simons Workshop on Mathematics and Physics. This work is supported by the Natural Sciences and Engineering Research Council of Canada.

\appendix

\section{Genericness of the near horizon extremal spinning (NHES) metric}\label{app:NHES}

Kunduri and Lucetti \cite{Kunduri:2008rs} have derived the most general 4-dimensional vacuum metric representing the geometry near an axisymmetric, extremal, and non-toriodal horizon:
\begin{equation}
	ds^{2}= \Xi(\zeta) \left[ -\frac{\bar{r}^{2}}{r_{0}^{2}} dv^{2} + 2\,dv\, d\bar{r} \right] + \frac{\Xi(\zeta)}{Q(\zeta)} d\zeta^{2} + \frac{Q(\zeta)}{\Xi(\zeta)}(dx+\bar{r}\,dv)^{2},
\end{equation}
with
\begin{equation}
	\Xi(\zeta) = \frac{1}{b} + \frac{b\zeta^{2}}{4}, \quad Q(\zeta) = \frac{1}{\lambda b^{2} r_{0}^{2}}\ (\lambda b\zeta+2)\left(2\lambda-b\zeta\right),
\end{equation}
where $(b,r_{0},\lambda)$ are constants.  Changing coordinates via
\begin{equation}
	v = \bar{t} - \frac{r_{0}^{2}}{\bar{r}}, \quad \zeta = \frac{\lambda^{2}-1+(\lambda^{2}+1)\cos\theta}{b\lambda}, \quad x = -r_{0}^{2}\left( \frac{b\lambda\phi}{\lambda^{2}+1} + \ln\frac{\bar{r}}{r_{0}}\right),
\end{equation}
puts the metric in the NHES form
\begin{subequations}
\begin{gather}
	ds^{2} = \Gamma(\theta) \left( -\frac{\bar{r}^{2}}{r_{0}^{2}} d\bar{t}^{2} +  \frac{r_{0}^{2}}{\bar{r}^{2}} dr^{2} + r_{0}^{2} d\theta^{2} \right) + \frac{\sin^{2}\theta}{\Gamma(\theta)}  \left( \frac{\gamma \bar{r}}{r_{0}} d\bar{t} -r_{0}d\phi \right)^{2}, \\
	\Gamma(\theta) = \tfrac{1}{2} \alpha(\cos^{2}\theta + 1) + \beta \cos\theta,
\end{gather}
\end{subequations}
with
\begin{equation}
	\alpha = \frac{(\lambda^{2}+1)^{2}}{2b\lambda^{2}}, \quad \beta = \frac{1-\lambda^{4}}{2b\lambda^{2}}, \quad \gamma^{2} = \alpha^{2} - \beta^{2}.
\end{equation}
Therefore, the NHES metric used in the main text is just another representation of the most general metric representing the near geometry around an axisymmetric-extremal horizon in 4-dimension vacuum general relativity.  We conclude by noting that the absence of conical singularities demands $\alpha = 1$ and $\beta = 0$ and hence reproduces the NHEK geometry; i.e., the only extremal near horizon vacuum geometry with $S^{2}$ topology in 4 dimensions is extremal Kerr, as first pointed out in \cite{Kunduri:2008rs}.

\bibliographystyle{JHEP}
\bibliography{TS_manuscript}

\end{document}